\documentclass[aps,prl,article,twocolumn,showpacs,preprintnumbers,amsmath,amssymb,superscriptaddress]{revtex4}

\date{\today}
\usepackage{epsfig}
\usepackage{subfigure}
\usepackage{graphicx}
\usepackage{dcolumn}
\usepackage{bm}
\usepackage[colorlinks,linkcolor=blue,hyperindex,CJKbookmarks]{hyperref}
\usepackage{float}
\usepackage{hyperref}
\usepackage{comment}
\begin{document}

\title{Real-Space Visualization of Remnant Mott Gap and Magnon Excitations}
\author{Y. Wang }
 \affiliation{Department of Applied Physics, Stanford University, California 94305, USA} \affiliation{SLAC National Accelerator Laboratory, Stanford Institute for Materials and Energy Sciences, 2575 Sand Hill Road,
Menlo Park, California 94025, USA}
\author{C. J. Jia}
\affiliation{Department of Applied Physics, Stanford University, California 94305, USA}
\affiliation{SLAC National Accelerator Laboratory, Stanford Institute for Materials and Energy Sciences, 2575 Sand Hill Road,
Menlo Park, California 94025, USA}
\author{B. Moritz}
\affiliation{SLAC National Accelerator Laboratory, Stanford Institute for Materials and Energy Sciences, 2575 Sand Hill Road,
Menlo Park, California 94025, USA}
\affiliation{Department of Physics and Astrophysics, University of North Dakota, Grand Forks, North Dakota 58202, USA}
\author{T. P. Devereaux}
 \email[Author to whom correspondence should be addressed to Y. W. (\href{mailto:yaowang@stanford.edu}{yaowang@stanford.edu}) or T. P. D. (\href{mailto:tpd@stanford.edu}{tpd@stanford.edu})
]{}
\affiliation{SLAC National Accelerator Laboratory, Stanford Institute for Materials and Energy Sciences, 2575 Sand Hill Road,
Menlo Park, California 94025, USA}
\affiliation{Geballe Laboratory for Advanced Materials, Stanford University, California 94305, USA}
\date{\today}
\begin{abstract}
We demonstrate the ability to visualize real-space dynamics of charge gap and magnon excitations in the Mott phase of the single-band Hubbard model and the remnants of these excitations with hole or electron doping.  At short times, the character of magnetic and charge excitations is maintained even for large doping away from the Mott and antiferromagnetic phases.  Doping influences both the real-space patterns and long timescales of these excitations with a clear carrier asymmetry attributable to particle-hole symmetry breaking in the underlying model.  Further, a rapidly-oscillating charge density wave-like pattern weakens, but persists as a visible demonstration of a sub-leading instability at half-filling which remains upon doping.  The results offer an approach to analyzing the behavior of systems where momentum-space is either inaccessible or poorly defined.
\end{abstract}

\maketitle

Demand for knowledge about inhomogeneous nano-scale structure in physics and chemistry places increased emphasis on real space imaging and spectroscopy. Coupled with developments in ultrafast laser techniques, observation of femto- and even atto-second dynamics has enhanced our understanding of chemical reactions and charge transport\cite{AutosecondMethod, AutosecondLaser}. This is particularly interesting in correlated systems where the electrons' spin and charge degrees of freedom are intimately intertwined\cite{Davis29102013}. A canonical example is the Mott gap excitation out of an antiferromagnetic (AFM) ground state, prevalent in many correlated oxides such as the cuprates, which reveals information about a competing ground state such as superconductivity or charge density order.
The frequency and momentum structure of Mott gap excitations have been identified by resonant inelastic X-ray scattering (RIXS) at the Cu $K$-edge\cite{MottGap2003, MottGap2005}; however, how these excitations evolve with doping and their real-space spin and charge structure and time scales are essential information. For example, do the characteristics of electron correlation, embodied in the large Coulomb repulsion $U$, become less relevant with doping due to screening as a weakly correlated Fermi liquid develops?

Inelastic x-ray scattering (IXS) provides a promising route to answer these questions.  It provides direct access to the charge dynamical structure factor $S^{(c)}({\bf{q}},\omega)$, and has been able to reveal time-dependent phonon correlations in Ge\cite{ReisNPhys} and the spatio-temporal landscape of charge excitations in water and graphene\cite{Peter_PRL,Peter_Science}, raising expectations that a similar approach could be used in correlated systems. Yet non-resonant IXS cannot capture spin dynamics.  However, with improvements in momentum and energy resolution, RIXS, exploiting spin-orbit coupling at the transition-metal $L$-edge, has been used to investigate magnetic excitations encoded in the spin dynamical structure factor  $S^{(s)}({\bf{q}},\omega)$ in recent studies\cite{PhysRevLett.104.077002, RevModPhys.83.705, le2011intense, PhysRevLett.102.027401, PhysRevLett.103.047401, PhysRevLett.102.167401, dean2013persistence,ChunjingLedge} complementary to traditional inelastic neutron scattering (INS)\cite{brockhouse1995slow}. This raises the possibility of visualizing the spin and charge nature of
Mott gap excitations and their doping evolution on the intrinsic length and time scales for collective electron behavior.

\begin{figure}[!t]
\centering
\includegraphics[width=7cm]{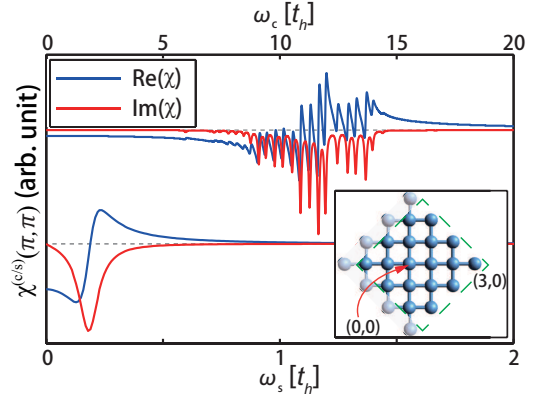}
\caption{\label{fig:1} Charge (top) and spin (bottom) density response as a function of frequency at $(\pi,\pi)$ for half-filling. The blue (red) lines represent the real (imaginary) parts of $\chi^{(c/s)}(\pi,\pi)$ and the dashed lines denote the baseline (or zero). The inset shows the 18A real-space Betts cluster employed in this study\cite{Betts:1999vf}.}
\end{figure}

\begin{figure*}[!ht]
\begin{center}
\includegraphics[width=16cm]{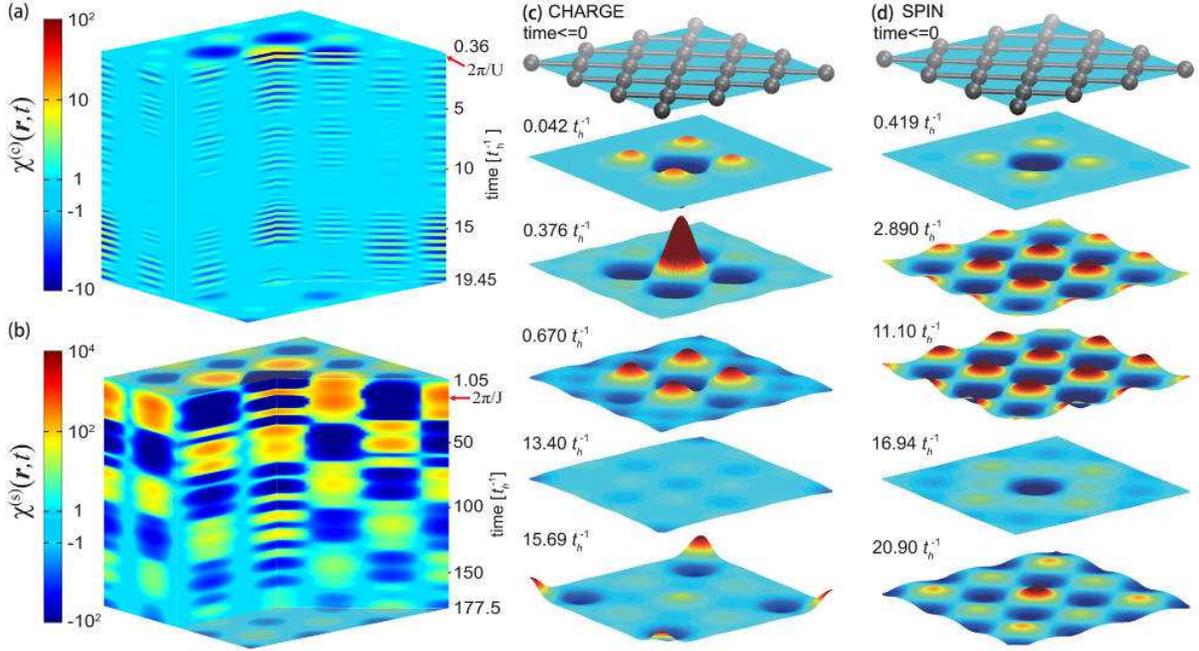}
\caption{\label{fig:2} Evolution of (a) charge and (b) spin susceptibility $\chi^{(c/s)}({\bf{r}},t)$ at half filling in the real-space cell. Color represents the intensities on a logarithmic scale.  Characteristic time scales for charge and spin response $2\pi/U$ and $2\pi/J$ are marked in the panels, respectively. (c,d) Featured frames of $\chi^{(c,s)}({\bf{r}},t)$ at representative times, respectively. The real-space cluster has been overlaid on the first frames to indicate the position of each site for clarity.}\end{center}
\end{figure*}

In this Letter, we evaluate the charge and spin dynamical structure factors $S^{(c/s)}(\textbf{r},t)$ and density response functions $\chi^{(c/s)}(\textbf{r},t)$\cite{Peter_Adv} of the single-band Hubbard model. We use exact diagonalization (ED) on small clusters to elucidate the real-space and time-dependent behavior of Mott gap and magnon excitations in correlated systems and their evolution with doping.  We find short-ranged, rapidly oscillating charge excitations and long-ranged, slower spin excitations reflecting disparate energy scales and damping for the excitations as well as the long-ranged nature of AFM at half-filling.  In doped systems remnant Mott gap and magnon excitations demonstrate the continued relevance of correlations, intrinsically different from the response expected for a weakly correlated fluid.  An observed charge density wave (CDW)-like pattern in the charge response reflects a sub-leading CDW instability at half-filling, which
competes with the dominant AFM order.  The lifetime and real-space differences which emerge with doping indicate that both charge and spin excitations are more well-defined in electron-doped rather than hole-doped systems.

An understanding of the low-energy physics of many correlated electron systems can be obtained by considering the single-band Hubbard model\cite{anderson1987resonating,Zhang:1988jf, Eskes:1988ef}:
\begin{equation}
H=-\sum_{i,j,\sigma} t_{ij}c^\dagger_{i\sigma}c_{j\sigma}+U\,\sum_i n_{i\uparrow}n_{i\downarrow},
\end{equation}
where $c^\dagger_{i\sigma} (c_{i\sigma})$ is the electron creation (annihilation) operator with spin $\sigma$ on site $i$, $t_{ij}$ is the hopping which is restricted to nearest ($t_{\langle ij\rangle}=t_h$) and next-nearest neighbors ($t_{\langle\langle ij\rangle\rangle}=t^\prime_h$), and $U$ controls the strength of the on-site electron-electron interaction.

To study the behavior of charge and spin excitations, we evaluate the response function $\chi(\textbf{k},\omega)$ through the fluctuation-dissipation theorem from $S(\textbf{q}, \omega)$, defined as
\begin{equation}\label{CFE}
S^{(\alpha)}(\textbf{q},\omega)=\frac1{\pi}\textrm{Im}\langle0|\rho^{(\alpha)}_{-\textbf{q}}\frac1{H-E_0-\omega-i\Gamma}\rho^{(\alpha)}_{\textbf{q}}|0\rangle,
\end{equation}
where $\alpha$ denotes either charge ($c$) or spin ($s$), with the density operator $\rho^{(c/s)}_{\textbf{q}}=\sum_{\textbf{k}}\left(c^\dagger_{\textbf{k}+\textbf{q},\uparrow}c_{\textbf{k},\uparrow}\pm c^\dagger_{\textbf{k}+\textbf{q},\downarrow}c_{\textbf{k},\downarrow}\right)$, respectively, $|0\rangle$ and $E_0$ are the ground state and energy. Final state lifetime effects, or line-widths, are approximated via an {\it ad hoc} Lorentzian broadening (HWHM) $\Gamma$. The parameters are set to be $U=10\,t_h$, $t_h^\prime=-0.25\,t_h$ and $\Gamma=0.05\,t_h$ in our calculations. For $t_h\sim 0.35$ eV the corresponding unit of time would be $t_h^{-1}\sim 2$ fs.

We use the parallel Arnoldi method\cite{lehoucq1998arpack} to determine the ground state wavefunction and continued fraction expansion\cite{Dagotto:1994cz} to obtain $S^{(c/s)}({\bf q},\omega)$\cite{altland2006condensed, ChunjingNJP} and $\chi^{(c/s)}({\bf q},\omega)$ (real and imaginary parts shown in Fig.~\ref{fig:1} for ${\bf q}=(\pi,\pi)$ at half-filling).  The calculations are performed on the 18A Betts cluster\cite{Betts:1999vf} with periodic boundary conditions (see inset of Fig.~\ref{fig:1}), chosen to strike a balance between momentum-space resolution/real-space cover and the Hilbert space dimension which increases exponentially with cluster size. 
A Fourier transform to real-space and time yields the charge (spin) response of the system $\delta n({\bf r},t)$ ($\delta m({\bf r},t)$) to unit perturbations $\delta({\bf r},t)$ at the center of the finite-sized cluster, the instantaneous introduction of a new charge carrier (spin) to the system.


Figs.~\ref{fig:2}(a) and (c) display the time evolution of $\chi^{(c)}({\bf{r}},t)$ at half-filling, while Figs.~\ref{fig:2}(b) and (d) show $\chi^{(s)}({\bf{r}},t)$, but over a different temporal range due to differences in the characteristic time scales.  Animations of the time evolution corresponding to a fine-grained sequence of frames like those shown in Figs.~\ref{fig:2}(c) and (d) can be found in the online SM.

The charge excitation patterns reflect evolution following instantaneous modification of the occupancy realized by exciting across the Mott gap with a characteristic oscillation period of $0.63\,t_h^{-1}\scriptsize{\sim}2\pi/U$. The evolution shows rapid oscillations and reflects the short-ranged, local screening of the introduced charge at the perturbed site.  The real-space pattern can be associated closely to a very local $(\pi,\pi)$ CDW-like pattern, which indicates that the propagation of the Mott gap excitation can be characterized by the temporal evolution or motion of a CDW-like state with short-ranged correlations. This is consistent with other studies that find a $(\pi,\pi)$ CDW as a sub-leading instability in the Hubbard model at half-filling\cite{PhysRevB.12.5249, PhysRevLett.53.2327,PhysRevLett.66.778}, that can be stabilized by considering the influence of electron-phonon
coupling\cite{BethPRL}.  This pattern rapidly decays, highlighting the damped nature of the Mott gap excitation and reflecting its rather broad structure in the frequency domain (see the effective width of the charge response in Fig.~\ref{fig:1}).

In contrast to Mott gap excitations, spin excitations reflected in $\chi^{(s)}({\bf{r}},t)$ are long-ranged and evolve on much longer time scales with a period of roughly $8\,t_h^{-1} \scriptsize{\sim} 2\pi/2J$ (see Figs.~\ref{fig:3}(b)), where $J \scriptsize{\sim} 4t_h^2/U$.  The AFM character of these excitations persists with an envelope whose period is roughly $32\,t_h^{-1} \scriptsize{\sim} 2\pi/(J/2)$ and an overall damping of approximately the same scale.  The distribution of changes to the spin intensity is more uniform, longer-ranged, and persistent than that for charge at half-filling which reflects the relative strength of both high energy magnons near the AFZB (the top of the magnon band at an energy $\sim$$2J$) and the dominant low energy magnons at larger momenta (the soft magnetic mode near $(\pi,\pi)$).  The damping and long-time envelope have similar time scales attributable to similar scales for the line-width and soft mode energy due to finite-size effects on the small cluster (see Fig.~\ref{fig:1} and the SM).

\begin{figure}[!ht]
\centering
\includegraphics[height=8.45cm]{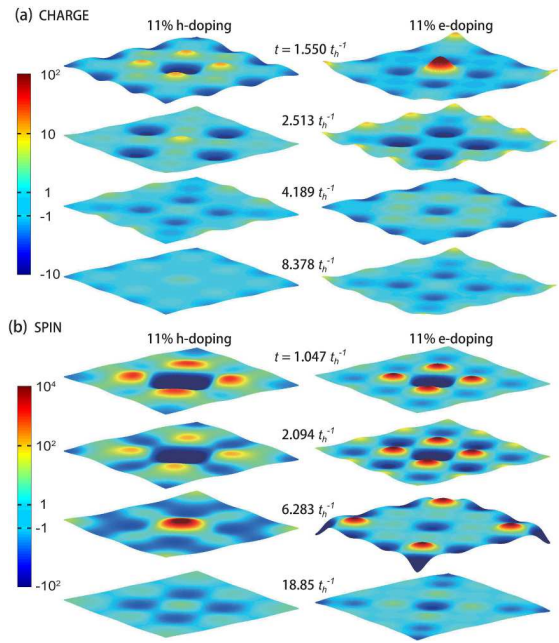}
\caption{\label{fig:4} (a) Comparison of $\chi^{(c)}({\bf{r}},t)$ for 11.1\% hole- (left) and electron-doping (right). (b) Comparison of $\chi^{(s)}({\bf{r}},t)$ for 11.1\% hole- (left) and electron-doping (right).}
\end{figure}

\begin{figure*}[!ht]
\centering
\includegraphics[height=5.95cm]{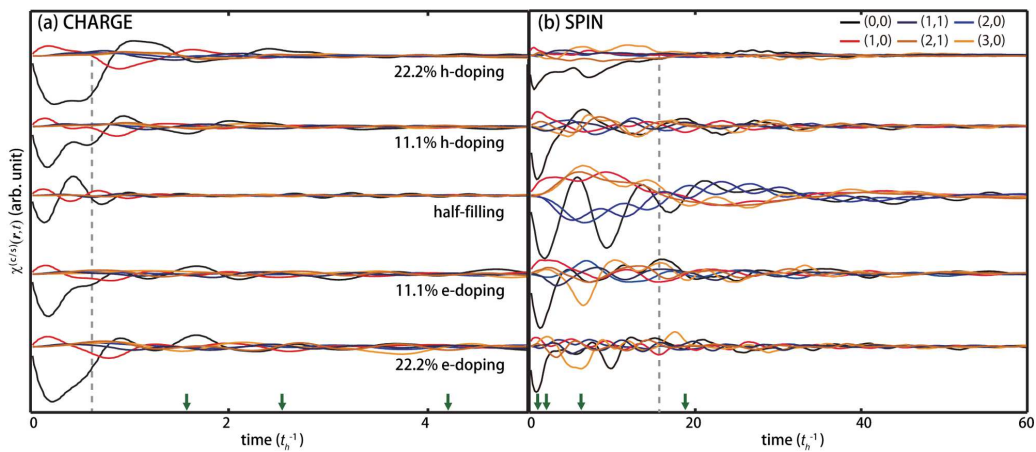}
\caption{\label{fig:3}(a) Charge and (b) spin density response functions $\chi^{(c/s)}({\bf{r}},t)$ at different real-space coordinates and doping levels. The various blue (red) shaded lines represent progressively longer range response functions on the same (opposite) AFM sublattice. The grey dashed line marks the characteristic time scale $2\pi/U$ ($2\pi/J$) for charge (spin). The green arrows highlight the corresponding time for each frame in Fig.~\ref{fig:4}.}
\end{figure*}

A clear view of the evolution of charge and spin excitations with doping can be seen in the corresponding real-space and time response functions shown in Figs.~\ref{fig:4} and ~\ref{fig:3} as well as more detail in Figs.~S2 and S3 in the SM. Comparing the charge response at different doping levels (see Figs.~\ref{fig:4}(a),~\ref{fig:3}(a), and S2), one observes the development of a slower component whose time scale increases with doping as the Mott gap closes and spectral weight transfers to lower frequencies.  Nevertheless, at short times, {\it e.g.} $\chi^{(c)}({\bf{r}},t)$ at $0\scriptsize{\sim} 0.3\,t_h^{-1}$, the response remains similar irrespective of doping. At longer times ($\gtrsim0.3\,t_h^{-1}$), the strength of the remnant Mott gap excitations decreases while the slower oscillations grow and dominate at large doping.  In addition, a comparison of the charge response between hole- and electron-doping indicates that excitations in electron-doped systems survive longer (to higher doping levels or for longer times at a similar doping) than those of hole-doped systems (see Figs.~\ref{fig:4}(a) and \ref{fig:3}(a) after $3\,t_h^{-1}$).

Observations of real-space temporal dynamics provide information about the doping dependent charge response in momentum and frequency.  While doping leads to a transfer of spectral weight to lower energies near the Fermi level (chemical potential), remnant Mott gap excitations persist at an energy scale set by $\sim U$ which indicates the continued relevance of electronic correlations with doping. The lower energy excitations possess narrower line-widths (longer lifetime) compared with those across the Mott gap.  Asymmetry of the dynamics between electron- and hole-doping reflects differences in spectral weight transfer between the UHB and LHB\cite{Brian_Raman}. The difference of lifetime also indicates broader (in terms of HWHM) excitations peaked at lower energy with hole doping (see Fig.~S1(a)). Physically it suggests a more well-defined quasi-particle with longer lifetime in electron-doped systems in agreement with other numerical techniques such as Monte Carlo\cite{Brian_Raman}.


Strong AFM patterns persist in the spin response with light doping (11.1\% hole or electron); however, this AFM pattern weakens significantly when heavily doped (22.2\%) with a clear electron-hole asymmetry (see Fig.~\ref{fig:3}(b)).  A distinctive AFM pattern does not appear for 22.2\% hole-doping, while for electron-doping this pattern partially remains, highlighting the comparative robustness of the AFM phase.  In terms of dynamics, both hole and electron doping lead to shorter oscillation periods and a decrease in intensity.  Spectral weight, originally peaked at $(\pi,\pi)$ in the half-filled system, crosses-over as a function of doping and becomes more peaked near zone center\cite{ChunjingLedge}. The overall reduction in the period and intensity reflects a harder spin response with a reduced magnitude and increased line-width.  This hardening can have two origins.  First, there is a natural cross-over from a low-energy, AFM response dominated by momentum $(\pi,\pi)$ to one in which the response at $(\pi,\pi)$ has been suppressed significantly and spectral weight transferred to persistent, high-energy AFM zone boundary spin excitations (or even smaller ${\bf q}$)\cite{Neutron_Wakimoto,Neutron_Fujita,le2011intense,Dean_RIXS,Minola_RIXS,ChunjingLedge}.  Second, these high-energy zone-boundary spin excitations themselves may harden as a function of doping
which appears to be the case with electron-doping as observed in the frequency domain\cite{WSLee_preprint,ChunjingLedge}.

To summarize, we calculated $S^{(c/s)}(\textbf{q},\omega)$ for the single-band Hubbard model on a small cluster using ED and determined $\chi^{(c/s)}(\textbf{x},t)$.  We observe the evolution of Mott gap excitations at half-filling with a CDW-like pattern and period set by $U$, which persist as remnant excitations with doping.  The spin response function at half-filling is dominated by well-defined, long-range AFM excitations with a more complicated temporal structure set by both high-energy magnons near the AFZB and the dominant soft magnetic mode near $(\pi,\pi)$.  Compared to half-filling, doping leads to increased charge response time scales for electron-doping compared to the same hole-doping level and the AFM character of the spin response is reduced more significantly with hole-doping than electron-doping. This behavior reflects underlying electron-hole asymmetry\cite{Brian_Raman,hanke2010book,anderson2011personal,anderson2006theory}.

In the study of Mott gap and magnon excitations, the interesting time scales are on the order of femtoseconds or longer, requiring information over only a relatively small energy range ($\sim$$10\,$eV). This differs from the requirements for IXS in water\cite{Peter_PRL} and Compton scattering in cuprates\cite{sakurai2011imaging}, which look at fundamentally shorter times associated with screening and orbital dynamics. However, with RIXS one may need to disentangle intra-atomic $dd$ excitations\cite{RevModPhys.83.705} from the Mott (or charge transfer) excitations which may occur on similar scales in transition metal (TM) oxides. At the other end of the spectrum, damping of these excitations sets the effective time window and limits the necessary energy resolution. With improvements in resolution, one may need to consider the influence of phonons on the low-energy spectrum.
For RIXS at TM $L$-edges, the best available resolution is $\sim$100\,meV\cite{PhysRevLett.104.077002, RevModPhys.83.705}.

Real space resolution is limited by the momentum transfer accessible in an experiment which depends on photon energy.   High momentum transfers (near ($\pi$,$\pi$)) are not accessible in 3$d$ TM compounds (Ti $L_3$-edge $\sim$450\,eV -- Cu $\sim$930\,eV)\cite{PhysRevLett.104.077002,PhysRevLett.TiO,ghiringhelli2005NiOL,ghiringhelli2006Mn}; however, the situation is better in 4$d$ and 5$d$ TMs (such as ruthanates $\sim$2.8\,keV and iridates $\sim$11.2\,keV\cite{kim2009phase,ishii2011Ir}). Therefore, under the right conditions, RIXS can be a powerful tool for studying the spatio-temporal Mott gap and magnon excitations.
This method also can be applied to predict and analyze dynamics for systems lacking translation symmetry, {\textit{e.g.}}~molecules and nanoscale open boundary systems with active sites based on TMs manifesting correlated excitations. Standing-wave techniques \cite{schulke1981compton,golovchenko1981coherent,abbamonte2009implicit,gan2012crystallographic} can be applied to obtain $\chi(\textbf{r}_1,\textbf{r}_2,t)$ and provide a real space and time map of reaction pathways in these systems.

We thank K. Wohlfeld for helpful discussions and F. Liu for part of picture design. This work was supported at SLAC and Stanford University by the US Department of Energy, Office of Basic Energy Sciences, Division of Materials Science and Engineering, under Contract No. DE-AC02-76SF00515 and by the Computational Materials and Chemical Sciences Network (CMCSN) under Contract No. DE-SC0007091. Y.W. and C.J.J. were also supported by the Stanford Graduate Fellows in Science and Engineering. A portion of the computational work was performed using the resources of the National Energy Research Scientific Computing Center supported by the US Department of Energy, Office of Science, under Contract No. DE-AC02-05CH11231.

\bibliography{paper}

\end{document}